\newif\iffigs\figsfalse
\figstrue

\iffigs
\input epsf 

\else

\fi
\def\bbbz{{\bf Z}}
\def\bbbc{{\bf C}}
\def\bbbr{{\rm I\!R}} 

\def\bbb1{{\rm 1\!1}}

\newcommand{\eqnl}[2]{\par\parbox{11cm}
{\begin{eqnarray*}#1\end{eqnarray*}}\hfill
\parbox{1cm}{\begin{eqnarray}\label{#2}\end{eqnarray}}\break}

\newcommand{\eqngrlb}[3]{\par\parbox{11cm}
{\begin{eqnarray}\fbox{$\displaystyle#1\\#2$}\end{eqnarray}}\hfill
\parbox{1cm}{\begin{eqnarray}\label{#3}\end{\eqnarray}}\break}

\newcommand{\eqngrl}[3]{\par\parbox{11cm}
{\begin{eqnarray*}#1\\#2\end{eqnarray*}}\hfill
\parbox{1cm}{\begin{eqnarray}\label{#3}\end{eqnarray}}\break}

\newcommand{\eqngrrl}[4]{\par\parbox{11cm}
{\begin{eqnarray*}#1\\#2\\#3\end{eqnarray*}}\hfill
\parbox{1cm}{\begin{eqnarray}\label{#4}\end{eqnarray}}\break}

\newcommand{\refs}[1]{(\ref{#1})}
\def\PSL{P\!S\!L}
\def\apendix{\par
\setcounter{section}{0}
\setcounter{subsection}{0}
\def\thesection{Appendix \Alph{section}.}
\def\theequation{\Alph{section}.\arabic{equation}}
}

\documentstyle[12pt]{article}

\advance\hoffset by -.35in
\begin{document}
\bibliographystyle{unsrt}
\begin{flushright}
DIAS-STP/97-06\\
\end{flushright}

\def\pr{\prime}
\def\pa{\partial}
\def\es{\!=\!}
\def\ha{{1\over 2}}
\def\>{\rangle}
\def\<{\langle}
\def\mtx#1{\quad\hbox{{#1}}\quad}
\def\pan{\par\noindent}
\def\lam{\lambda}
\def\La{\Lambda}

\def\A{{\cal A}}
\def\G{\Gamma}
\def\Ga{\Gamma}
\def\F{{\cal F}}
\def\J{{\cal J}}
\def\M{{\cal M}}
\def\R{{\cal R}}
\def\W{{\cal W}}
\def\tr{\hbox{tr}}
\def\al{\alpha}
\def\d{\hbox{d}}
\def\De{\Delta}
\def\L{{\cal L}}
\def\H{{\cal H}}
\def\Tr{\hbox{Tr}}
\def\I{\hbox{Im}}
\def\R{\hbox{Re}}
\def\ti{\int\d^2\theta}
\def\bti{\int\d^2\bar\theta}
\def\ttbi{\int\d^2\theta\d^2\bar\theta}
\font \smallescriptfont       = cmr5 at 7pt

\begin{center}
\Large{{{\bf On the Uniqueness of the effective Lagrangian for 
{\bf N= 2 SQCD}}}}
\end{center}

\begin{center} M. Magro, L.O'Raifeartaigh and I. Sachs\\
{\it Dublin Institute for Advanced Studies,\\
10 Burlington Road, Dublin 4, Ireland.}
\end{center}

\begin{abstract}
The low energy effective Lagrangian for $N\es 2$ $SU(2)$ supersymmetric 
Yang-Mills theory coupled to $N_F<4$ massless matter fields is derived 
from the BPS mass formula using asymptotic freedom and 
assuming that the number of strong coupling singularities is finite. 

\end{abstract}

\section{Introduction} 
Over the last three years important progress has been made 
on the problem of strongly coupled Yang-Mills theory with extended 
supersymmetry, pioneered by the work of Seiberg and Witten \cite{SW1,SW2} for 
an $SU(2)$ gauge group. 
Assuming duality, in the sense that for strong coupling the low energy 
effective action is dominated 
by magnetic monopoles, they obtained an exact result for the low energy 
effective Lagrangian. Extensions to higher 
groups were later proposed \cite{Lerche}. The assumption of the 
monopole dominance is self-consistent and well motivated on physical grounds.
Nevertheless one expects this to be a consequence 
of rather more fundamental properties of the theory. Apart from clarifying 
the underlying structure of the solutions obtained so far investigation of 
this problem should be particularly interesting in view of possible generalizations. 

In the case of $N\es2$ Yang-Mills theory without matter results in this 
direction have been reported previously. In particular 
the Seiberg-Witten solution \cite{SW1} was re-derived in \cite{Matone2} 
under the assumption that the moduli space of inequivalent vacua was of 
genus zero and that it was parameterized by the Seiberg-Witten parameter $u$. 
More recently \cite{USW1} it has been shown 
neither of these assumptions is necessary. The 
same result can already be derived from the BPS-mass formula 
and the assumption that the moduli 
space is simply the space of inequivalent (complex) couplings $\tau$. 
The central ingredient in this analysis is the so-called 
{\it maximal equivalence group} ${\cal{G}}$ whose elements relate 
equivalent vacua. 
The equivalence is with respect to the mass-spectrum of the full tower of 
BPS-saturated states 
(not the subset of {\it stable} BPS-states). 
That ${\cal{G}}$ is not trivial follows on general grounds 
from the periodicity of the $\theta$-vacuum on one hand and the anomalous 
electric charge of the monopoles induced by the $\theta$-angle on the 
other. At the same time 
${\cal{G}} \subset U(1)\times P\!S\!L(2,{\bf Z})$ due to 
the particular form of the BPS-formula \cite{SW1}.

In this paper we consider the case where the 
$N\es2$ YM-theory is coupled to $N_F \!<\!4$ massless 
matter hypermultiplets \cite{SW2} in the fundamental representation of 
$SU(2)$. $N_F\!<\!4$ is required by asymptotic freedom which we shall use 
throughout. The moduli 
space is parameterized by a possibly multiple covering of the 
fundamental domain of ${\cal{G}}$ in the $\tau$-plane. Thus $\tau$ is not 
necessarily a uniformizing parameter. 
A further ingredient for our strong-coupling analysis is the asymptotic 
freedom which requires that the mass of the lightest charged field be 
finite in the non-perturbative regime. It turns out that the $S\!L(2,{\bf Z})$ 
property and the finite mass constraint are almost, but not quite strong 
enough to determine the low-energy effective action uniquely. Further 
constraints are then obtained by adding a further observable to the system. 
This observable is provided by the 
superconformal anomaly \cite{WH1}. We then show that the finite mass 
constraint \cite{USW1} plus the 
condition that the anomaly be a $U(1)$-section under ${\cal{G}}$ determine 
the set of possible fundamental domains for the coupling $\tau$.  
For $N_F\es 0,1,2$ there is a unique domain, which corresponds to the 
S-W-solution for these cases. For $N_F\es3$ there are several  
different domains, but only for one of them is the superconformal 
anomaly automatically a $U(1)$-section. The corresponding solution is 
precisely  the S-W-solution. For all the other domains the conditions that the 
anomaly be a $U(1)$-section leads to a formally overdetermined system. 
Whether all of them are actually overdetermined cannot be decided from 
the existing mathematical literature. It is not even clear whether 
these domains correspond to equivalence groups. 

The role played here by the superconformal anomaly is reminiscent of that 
played by conformal- or axial anomalies to obtain exact results in 
$2$-dimensional models. The superconformal anomaly has previously been used 
to compute certain $N\es2$ supersymmetric Green functions \cite{HW2}. Here we 
show that, in conjunction with the 
$S\!L(2,{\bf Z})$-structure, the superconformal anomaly 
determines the low-energy effective Lagrangian and completely parametrises 
the quantum moduli space. What distinguishes the $4$-dimensional model is 
that the non-perturbative contributions are not computable by other means 
(except for $1$ and $2$ instantons contributions   
\cite{Pouliot,Dorey,Ito,Aoyama,Harano}). 

The plan of this paper is as follows. In section \ref{sec2} we 
review the properties of $N=2$ Yang-Mills theory coupled 
to matter. In section \ref{sec3} we establish the necessary ingredients 
for the non-perturbative analysis in the later sections and recall the general 
construction of solutions \cite{USW1}. In section \ref{sec4} 
we give a quick derivation of the anomalous superconformal 
Ward-identity and obtain the constraints 
it imposes on the general solution. In section \ref{sec5} we 
obtain the unique solutions for $N_F\es 0,1$ and $2$, the SW-solution 
for $N_F=3$ and discuss the over determination problem for the other possible candidates. For the purpose of comparison with other papers \cite{Bilal,Ito-Yang} 
it is shown in the appendix 
how to obtain the second order differential equations  
satisfied by our solutions.

\section{Review of $N\es2$ YM-Theory with Matter}
\label{sec2}
We start with a review of $N\es2$ YM-theory with massless hypermultiplets in 
the fundamental representation. In addition to the canonical $N\es1$ 
kinetic terms and minimal gauge coupling for all the fields, 
$N\es2$-supersymmetry requires a superpotential
\eqnl{W=\sqrt{2}\tilde M_iAM^i,}{hp}
where $A$ is the chiral multiplet in the $N\es2$ Yang-Mills multiplet and 
$M_i;\;\;i\es1,\cdots, N_F$ are the matter hypermultiplets. 
The flat 
directions in the  potential for the scalar component $\phi$ of $A$ 
survive in the presence of hypermultiplets. More precisely, 
the potential vanishes for constant $\phi$ taking its value in the 
Cartan subalgebra of the gauge group $G$. In what follows we take 
$G\es SU(2)$. For $\phi\neq 0$, the Higgs 
mechanism breaks the gauge symmetry spontaneously to 
$U(1)$, inducing a mass term for all charged fields including the 
hypermultiplets as follows from \refs{hp}. The theory 
is then in the Coulomb branch \cite{SW2}. 
In this regime the most general low energy effective Lagrangian 
has a $F$-term of the form \cite{Sei,AP}
\eqnl{\G[\A]=\frac{1}{4\pi}\hbox{Im}\int d^4x
d^2\theta_1 d^2\theta_2 \; \F(\A),}{eff1}
where the prepotential $\F$, to be determined, is the result 
of integrating out the massive fields. 
In $N\es1$ notation \refs{eff1} becomes 
\eqnl{\G[A,W_\al]=\frac{1}{4\pi}\I\,\Tr\int\d^4x\Big\{ \ttbi 
\; (A_D\bar A-{\bar A}_D A)+\frac{1}{4}\ti\;\tau(A)\;W^\al W_\al\Big\},}{action2}
where $A=\phi+\dots$ and $W_\al$ are the chiral- and vector 
$N\es1$ superfields 
respectively. Furthermore 
\eqnl{ A_D=\ha\F'(A)\mtx{and} \tau(A)=\ha\F''(A)=A_D'(A).}{ada} 
The normalisations for $N_F\es0$ \cite{USW1} and $N_F>0$ \refs{ada} 
differ by a factor of $2$. 
Throughout this paper we take $N_F>0$ unless explicitly stated otherwise.  
Since $\tau(\phi)$ is the coefficient of the kinetic term in 
\refs{action2} its 
imaginary part must be positive. The real part of $\tau$ plays the 
role of an effective 
$\theta$-angle: $\mbox{Re }\tau = \frac{\theta}{\pi}$. Thus a shift of 
$\theta$ by $2 \pi$ corresponds to 
$\tau \mapsto T^2(\tau) = \tau +2$. 
Consequently, the observables of the low energy effective theory are 
invariant under $\{\tau\mapsto T^{2n}\tau, n \in \bbbz\}$. This is 
therefore the case, in particular for the mass of BPS-saturated 
states which, for consistency with  the SUSY-algebra, must be proportional to the 
central charge \cite{OW}
\eqnl{M=\sqrt{2}\vert Z\vert .}{Z0}
At the classical level the theory 
is parameterized by the two real parameters $|\phi^3|$ and $\I(\tau)$. The moduli space of inequivalent vacua is therefore $2$-dimensional. The central charge $Z$ is given by \cite{OW}
\eqnl{Z = \vert \phi^3 \vert \vert n_e + \tau n_m \vert}{Zcla}
where the integers $n_e$ and $n_m$ label electric and magnetic 
charges respectively.
In the quantum theory the mass of the lightest charged field $m$ 
sets the scale for the low energy coupling. For $N_F<4$ the theory is 
asymptotically free and hence 
perturbation theory is valid as long as $m\!>\!>\!\Lambda$. 
In particular at the semiclassical level we then have ($\Lambda\es 1$) 
\cite{Shif0} 
\eqnl{\tau(a)= \frac{i}{\pi} (4-N_F)\log (a) +c\mtx{where}a=
\ha \<\phi^3\> }{tau}
and $c$ depends on the renormalisation scheme adopted. 
Higher loop perturbative corrections to the running coupling are 
absent \cite{W2}. As explained in \cite{OW,Rocek} the topological nature of 
the mass spectrum guarantees that the structure for the BPS-spectrum is the 
same at  the quantum level. This implies in particular the linearity in 
$(n_e,n_m)$. The coefficients, however, can be modified. 
Computing the BPS-bounds from the gauge-invariant extension 
$\F(\sqrt{\A^a\A^a})$ of the low energy {\it effective} action ,  
it has been argued in \cite{SW1,Rocek}, that the quantum corrected 
BPS-formula is given by
\eqnl{ Z=an_e+a_Dn_m,}{Z}
where $a_D = \ha \F'(a)$. Alternatively one can compute the centre of the 
low-energy effective theory from
\eqnl{\{Q^{\hbox{{\smallescriptfont eff}}},
\bar Q^{\hbox{{\smallescriptfont eff}}}\}=i(P\!\!\!/^{\hbox{{
\smallescriptfont eff}}}-Z^{\hbox{{\smallescriptfont eff}}}),}{Ze1}
where $Q^{\hbox{{\smallescriptfont eff}}}_\al$ are the supercharges 
obtained from the low energy 
effective Lagrangian $\F(\sqrt{\A^a\A^a})$. One has \cite{MRSZ}
\eqngrl{Q^{\hbox{{\smallescriptfont eff}}}&=&\frac{1}{4\pi}\int d^3x(X+iY)^a[\I(\F_{ab})]^{-1}\psi^b +O(\psi^2),}
{\bar Q^{\hbox{{\smallescriptfont eff}}}&=&\pi_\psi^{T\;a}(X+iY)^{\dagger\;a}+O(\psi^2),}{qz1}
where $O(\psi^2)$ corresponds to terms in the Hamiltonian which vanish when the fermion fields are set to zero, and
\eqngrrl{X^a&=&\gamma^i\gamma^5 B_i^a-i4\pi\gamma^i[\I({\cal{F}}_{ab})]^{-1}
\tilde\pi^b_i+((\ha))[\I({\cal{F}}_{ab})]^{-1}{\cal{C}}^b\gamma^0\gamma^5,}
{Y^a&=&[\I({\cal{F}}_{ab})]^{-1}\pi_{\Phi}^b+\gamma^i\gamma^0D_i\Phi.\mtx{where}}
{\Phi^a&=&\R\phi^a+i\gamma^5\I\phi^a,\qquad \tilde\pi^a_i=
\frac{1}{4\pi}[\I(\F_{ab})]E^b}{qz2}
and ${\cal{C}}^a\es[\phi_D,\phi^\dagger]^a$. Computing the Poisson brackets 
of the effective supercharges one then again 
finds the result \refs{Z}. This shows that independent  of the actual form of 
the function ${\cal{F}}$ appearing in the effective action, the centre is given by 
\refs{Z}. Comparison with the results in \cite{Rocek} furthermore shows that 
the mass of the BPS-states indeed saturates the inequality $M\geq|Z|$ also at 
the quantum level, as anticipated in \cite{OW,SW1}. The details of this 
calculation will be presented 
elsewhere \cite{MRSZ}. To summarize, the low energy effective theory is 
parameterised by either of the complex parameters 
$a$ with values 
in $\bbbc$, or $\tau$ which takes any value in the upper half plane 
$H$. In particular, the space of inequivalent vacua, or moduli space, 
$\cal M$ is $1$-complex dimensional. 

\section{Non-Perturbative Contributions}
\label{sec3}
In addition to the perturbative corrections, reviewed in the last section, 
the low-energy effective action receives non-perturbative corrections due  
to topologically non-trivial contributions. The problem is then to 
determine $\F(a)$ or equivalently $\tau(a)$. 
We formulate the problem such as to 
make maximal use of the analyticity of $\tau(a)$ (which 
reflects the chirality of the supersymmetry algebra if derivatives are 
neglected) and the underlying 
$S\!L(2,{\bf Z})$-structure. 
Specifically our analysis uses the following properties:
\vskip 0.2truecm 
\noindent (a) The effective coupling constant  
$\tau=\frac{\theta_{\hbox{{\smallescriptfont eff}}}}{\pi}+i8\pi g_{\hbox{{\smallescriptfont eff}}}^{-2}$ takes all 
values in the upper half plane $H$.
\vskip 0.2truecm 
\noindent (b) The BPS-spectrum $M$ is a single valued function on the moduli 
space,  $M=M(P)$, $P\in {\cal M}$.\pan
\vskip 0.2truecm
\noindent (c) The mass $m$ of the lightest charged field 
(possibly composite) is finite except in the asymptotically free region.
\vskip 0.2truecm
\noindent (d) The superconformal anomaly $u$ is a $U(1)$ section under 
transformations of the equivalence group ${\cal{G}}$.  \pan
\vskip 0.2truecm 
\noindent (e) The set of singular points of ${\cal M}$ is finite.
\vskip 0.2truecm 
The condition (c) reflects 
the asymptotic freedom of the underlying non-abelian theory. As for (d) 
this condition will become clear in section \ref{sec4}. 
Finally, the condition (e) is a technical assumption, which might not be 
necessary.

At the quantum level the quantities $a$, $a_D$ and $\tau$ will be 
transcendental functions of each other. This can be seen already from the 
$1$-loop correction \refs{tau}. Correspondingly these functions have 
non-trivial Riemann surfaces. This multivaluedness leads then to 
identifications in the space of vacua, that is, certain points in the  
space of vacua are physically equivalent. To reduce this degeneracy we 
introduce the {\it maximal equivalence group} 
${\cal{G}}$ defined by its (linear) 
action on the vector ${\bf a}\es(a_D,a)$. The elements of ${\cal{G}}$ are the 
transformations which identify all different ${\bf a}$ corresponding 
to equivalent physical vacua 
$P$. Invariance of the BPS-spectrum $M$ then implies that\footnote{Up to the 
extra $U(1)$-factor the same condition was obtained in \cite{SW1} 
(see paragraph below eqn. (4.7)).} 
${\cal{G}}\subset U(1)\times\PSL(2,\bbbz)$. On the other hand 
the conditions $\I\tau \geq0$ and 
\eqnl{\tau(a)=\frac{\d a_D}{\d a},}{tau2}
imply that ${\cal{G}}$ is represented on $\tau$ by a subgroup $G$ of 
$\PSL(2,{\bf Z})$, the projective modular transformations. ${\cal{G}}$ is not trivial 
since $T^2\in G$. 
Note that the definition of the maximal 
equivalence group given here is slightly more general than that of 
\cite{USW1} for pure YM-theory. As we will see below, this more general 
setting does not lead to new solution in that case. \par
The moduli space is then in $1\!-\!1$ 
correspondence with a possibly multiple covering $D$ of $H/G$. 
We can therefore parameterise 
${\cal{M}}$ by the upper half plane $H$ by means of the Fuchsian map
\eqngrl{\tau:  H&\rightarrow& D}
{ z&\mapsto&\tau(z)=\frac{y_1(z)}{y_2(z)}\mtx{where}{\bf y}''+Q{\bf y}=0 }{ty}
and $2Q(z)=\{\tau,z\}$ is the Schwarzian of $\tau$. 
While the function $\tau(z)$ is normally complicated, $Q(z)$ 
has the simple form \cite{Nehari}
\eqnl{Q(z) = \ha\sum\limits_{i=1}^n\big[\ha\frac{1-\al_i^2}
{(z-a_i)^2}+\frac{\beta_i}{z-a_i}\big],}{q}
where $n+1$ is the number of edges of $D$, the 
$a_i$'s are the points on the real axis into which the 
corners of the polygon $D$ are mapped and the $\pi \alpha_i\in[0,\pi)$ 
the interior angles of $D$. Little is known about the geometrical 
interpretation of the {\it accessory parameters} 
$\beta_i$ for a general polygon \cite{Nehari,Venkov}. This is the origin of 
the 'technical problem' for $N_F\es3$ anticipated in the introduction. 
In \refs{q} we have chosen to map 
the weak-coupling singularity $\tau = i\infty$ to infinity in the 
$z$-plane. We then have \cite{Nehari}
\eqnl{Q(z) \simeq\frac{1}{4z^2}\mtx{for} z\rightarrow\infty.}{compact} 
For later use we define the index 
$\mu$ of $D$ by the ratio 
\eqnl{\mu=\frac{\hbox{Area}(D)}{\hbox{Area}(D_0)},}{mu}
where the area is defined with respect to the Poincar\'e metric 
\cite{Frakas,USW1} on the $\tau$ plane and $D_0$ is the fundamental domain of 
$P\!S\!L(2,{\bf Z})$. The index is then related to the angles $\alpha_i$
of $D$ by  
\eqnl{\sum_i(1-\alpha_i) = \frac{\mu}{3}+1.}{index}

As explained in \cite{USW1}, for a given equivalence group ${\cal{G}}$ 
the general solution for the section ${\bf a}$ is given by 
\eqnl{{{\bf a}}=f'{{\bf y}}-f{{\bf y}}'={{\bf W}}(f,{{\bf y}}),}{fy}
where $f$ is a $U(1)$-section under equivalence 
transformations, in order to leave the mass spectrum invariant,
${\bf y}$ is as in \refs{ty} and ${\bf W}(f,{{\bf y}})$ 
is the Wronskian of $f$ and ${\bf y}$. To complete the construction we need 
to match the boundary conditions 
with those given by the semiclassical contribution \refs{tau}. 
It follows from \refs{compact} and \refs{ty} that
\eqnl{{\bf y}(z)\simeq z^{{1 \over 2}}(c \, \hbox{ln}(z),1) 
\mtx{for} z\rightarrow \infty.}{assy2}
The constant $c$ is constrained by the $\theta$-vacuum symmetry 
explained above. In the presence of matter multiplets this symmetry is 
in fact enhanced. This is due to the absence of odd-instanton 
contributions \cite{SW2} enlarging the minimal symmetry 
$\theta\rightarrow \theta+2\pi$ to 
$\theta\rightarrow \theta+\pi$ or equivalently 
$\tau\rightarrow T(\tau)=\tau+1$. 
Therefore we identify $\tau$ and $\tau+1$ and hence 
\eqnl{\tau(z)\rightarrow {i\over \pi}\hbox{ln}(z)
\mtx{for} z\rightarrow\infty.}{asstau}
Comparing \refs{asstau} with the semiclassical result 
\refs{tau} we then obtain for $z\rightarrow \infty$
\eqnl{f(z)\simeq f_0\; z^{(6-N_F)/2(4-N_F)}.}{fg2}
This completes the construction of $a$ and $a_D$ for a given equivalence 
group. Without further constraints the non-perturbative contributions are 
then not uniquely determined. Indeed any Fuchsian function $\tau(z)$ 
mapping $H$ into a 
fundamental domain of $G$ and any $U(1)$-section $f(z)$ satisfying the 
boundary condition \refs{fg2} is a solution. This freedom 
is partly removed by the finite mass condition (c). As explained in 
\cite{USW1} this constraint leads to a lower bound for the exponents 
$r_z$ of $f$ at its singularities. More precisely 
\eqnl{\cases{r_{z} =\ha(1-\al_i)\mtx{or} 
r_{z} \geq\ha(1+\al_i) \mtx{for} z = a_i\cr
r_{z}\geq1\mtx{for}z\neq a_i}.}{fmc}
From \refs{fmc} we obtain in particular the necessary condition
\eqnl{\cases{r_{z} \geq\ha(1-\al_i) \mtx{for} z= a_i \cr
r_{z}\geq1\mtx{for}z\neq a_i}.}{nec}
On the other hand, in order to be well defined on the fundamental domain $D$ 
the exponents $r_z$ must satisfy the total residue 
condition \cite{USW1}
\begin{equation}
\sum\limits_{z{\rm \ interior\hspace{.38ex}}}r_{z}+\frac{1}{2}
\sum\limits_{z\in\bbbr}r_{z}+\ha r_{\infty }=0.\label{cc}
\end{equation}
Substituting the 
lower bound \refs{nec} into \refs{cc} and using \refs{index} 
and \refs{fg2} we find that all the singularities of $f$ coincide with 
those of 
$Q$ and that the area of $D$ is bounded by
\eqnl{\mu \leq\frac{6}{4-N_F}.}{bm}
Repeating the same analysis for $N_F\es0$ leads to \cite{USW1}
\eqnl{\mu \leq 3.}{bm0}
To summarize, the finite mass condition, which reflects the asymptotic 
freedom of the underlying non-abelian theory considerably reduces the set of 
admissible equivalence groups. However, a finite set of polygons 
remains and each of them is a candidate for a possible solution of the low 
energy effective action. In order to impose further constraints we need 
to extend the set of observables of the low energy theory. The superconformal 
anomaly introduced in the next section provides this observable.

\section{The Superconformal Anomaly}
\label{sec4}
\def\be{\beta}
\def\al{\alpha}
\def\dal{{\dot\alpha}}
\def\ti{\int\d^2\theta}
\def\ddti{\int\d^4x\d^2\theta}
\def\bti{\int\d^2\bar\theta}
\def\ttbi{\int\d^2\theta\d^2\bar\theta}
\def\ddttbi{\int\;\d^4x\d^2\theta\d^2\bar\theta}

\def\tbi{\int\d^2\bar\theta}
\def\tbti{\int\d^2\theta\d^2\bar\theta}
\def\cd{{\cal{D}}}
\def\de{\delta}

The BPS-formula used in the previous section relates $a$ and $\F'(a)$ to a 
physical observable. The invariance of the BPS-spectrum under equivalence 
transformations is responsible for the $\PSL(2,\bbbz)$-structre which played 
an important role in determining the most general equivalence group in the 
previous sections.  
The purpose of this section is to relate yet another physical object to $a$ 
and $\F(a)$. This will be the superconformal anomaly. The invariance of the 
anomaly will then be used  to further constrain the equivalence group. 

\subsection{The Superconformal Ward-Identity}
Following \cite{WH1} we first obtain a relation between the low energy 
effective action and the superconformal anomaly. The Ward identity is most 
easily derived in $N\es1$ superspace. 
Using the invariance of the classical action the 
Schwinger functional satisfies the formal identity for an arbitrary 
superconformal transformation 
\eqnl{\int\;d\mu(a,v)\;e^{S[H_{\al\dal},a,w_\al] +(J,a+\delta a)}
=\int\;d\mu(a,v)\;e^{S[H_{\al\dal}+\de H_{\al\dal},a,w_\al] +(J,a)}}{scw1}
where $a,v$ are the $N\es1$ integration variables and $H_{\al\dal}$ is the 
supergravity prepotential coupling to the supercurrent $j_{\al\dal}$. 
Expanding 
both sides of the equality to the first order and using 
\eqnl{J= \frac{\de\Gamma[A]}{\de A},}{scw2}
where $\Gamma[A]$ is the effective action for the chiral superfield $A$, 
we obtain
\eqnl{\int {\d}^4x{\d}^2\theta{\d}^2\bar\theta\;
\frac{\de\Gamma[A]}{\de A}\<\de A\>= \int {\d}^4x{\d}^2\theta{\d}^2\bar\theta\;
\de H^{\al\dal}\<j_{\al\dal}\>.}{scw3}
The $N\es1$ fields and the 
supergravity prepotential transform under an arbitrary transformation as 
\eqnl{\de A= \bar D^2L^\al D_\al A -q(\bar D^2 D_\al L^\al)A
\mtx{and} \de H^{\al\dal}
=(D^\al L^{\dal}-D^{\dal}L^\al),}{scw4}
respectively, where $L^\al$ is the parameter superfield \cite{RS} and 
$q$ is the $R$-weight of $A$. 
We now consider a global 
axial transformation $\bar D^2D_\al L^\al=i\Delta$, where $\Delta$ is a 
real parameter. From \refs{scw4} and \refs{action2} we then obtain for the 
variation of the low energy effective action
\eqnl{\de_\Delta\Gamma[A]=-\frac{iq}{4\pi}\Delta\ddttbi\;
\bigl(\F''(A)A\bar A -\F'(A)\bar A +\hbox{h.c.}\bigr)}
{scw4.5}
On the other hand we have from (\ref{scw3},\ref{scw4})
\eqngrl{\de_\Delta\Gamma[A]&=&\ddttbi\;
\bigl(L^\al D^\dal-L^\dal D^\al\bigr)j_{\al\dal}}
{&=&-2\Delta\bigl(\ddti\;S-\hbox{h.c.}\bigr),}{scw5}
where the last equality uses the anomaly equation \cite{CL,GW}
\eqnl{D^\dal j_{\al\dal}=D_\al S,}{scw7}
with $S$ a chiral superfield ($D_\dal S\es 0$). Rewriting the result 
\refs{scw5} in $N\es2$ formulation we then get
\eqnl{\R\int\;d^4x\d^4\theta\;\Bigl(\F(\A)-\ha\F'(\A)\A\Bigr)
=c\R\int\;\d^4x\d^4\theta\;i{\cal{U}},}{scw8}
where $c$ is a constant and ${\cal{U}}$ is the $N\es2$ anomaly multiplet. 
This is the sought relation between the prepotential and 
the super conformal anomaly ${\cal{U}}$.  The same relation has been derived 
in \cite{Matone} between $\F(a)$ and the Seiberg-Witten parameter $u$, 
assuming the Seiberg-Witten solution, for pure YM-theory. The derivation 
of the relation \refs{scw8} is similar to that in \cite{WH1}. It 
does not make any assumption on the form of 
$\F$ and furthermore establishes the physical nature of the Seiberg-Witten 
parameter $u$. 

\subsection{Constraints on the Accessory Parameters}
For our purpose the precise form of the anomaly is not important. 
We will only make use of the fact that $u\es{\cal{U}}\vert_{\theta=0}$ is a 
low energy observable and 
therefore a $U(1)$-section under equivalence transformations (the phase 
of $u$ is not observable as it can be absorbed in a redefinition of 
the superspace coordinate\footnote{This statement is equivalent to the fact 
that in $N\es 1$-language the anomaly is of the form 
\Tr $\bar A A+c\tilde F F$ which is clearly independent of the phase of 
$\phi$.} $\theta_\al$. 
%
%
In the perturbative limit 
we have from \refs{tau},\refs{compact} and \refs{asstau}
\eqnl{ u(z)\propto z^{\frac{2}{4-N_F}},}{asu}
where we have used $u\propto a^2$ in that limit. Now, using 
\eqnl{\frac{du}{dz}=ci\frac{d}{dz}\Bigl(\F(a)-\ha\F'(a)a\Bigr)
=ci  {\cal{W}}(a_D,a)_z,}{us1.5}
where ${\cal{W}}(a_D,a)_z$ is the Wronskian of $a_D$ and $a$, 
we obtain from \refs{fy}
\eqnl{\frac{du}{dz}=\frac{4}{4-N_F}f^2(\frac{f''}{f}+Q).}{us2}
The form of \refs{us2} suggests the Ansatz 
\eqnl{u(z)=\frac{4}{4-N_F} f^2(z)h(z)}{au} 
where $h$ is a $U(1)$-section.
The boundary conditions \refs{compact} and \refs{fg2} require for $h$ 
the asymptotic behaviour 
\eqnl{h(z) \simeq \ha\frac{1}{4-N_F}\frac{1}{z} \mtx{for large} z.}{at} 
Substitution of 
\refs{au} into \refs{us2} then leads to
\eqnl{2\frac{f'}{f}h +h' = \frac{f''}{f}+Q\equiv \tilde Q.}{tq}
By construction $\tilde Q$ has at most simple and double poles. At a double 
pole $a_i$, say of $\tilde Q$, $h$ then has a simple pole\footnote{
Although a second order pole would at first sight be compatible with the 
boundary condition \refs{fg2} for $N_F\es3$, explicit inspection of all 
possible polygons shows that this possibility is in fact excluded}. 
The residue $h_i$ is then obtained from \refs{us2}, i.e.
\eqnl{h_i=\ha\frac{r_i^2-r_i+\frac{1}{4}(1-\al_i^2)}{r_i-\ha}.}{bbq}
The residue $q_i$ of $\tilde Q$ at this 
point is then given in turn by 
\eqnl{q_i=2 r_i \Bigl(h(z)-\frac{h_i}{z-a_i}\Bigr)\vert_{z=a_i}
+ 2 h_i \Bigl(\frac{f'(z)}{f(z)}-\frac{r_i}{z-a_i}\Bigr)\vert_{z=a_i}.}{qt}
The only other possibility for $h$ to be singular is if $\tilde Q$ has 
a simple pole at $a_i$; this happens if and only if $r_i\es\ha(1\pm\al_i)$. 
Cancellation of the double poles on the left hand side of \refs{tq} then 
requires that $\al_i\es 0$. If $\al_i\neq 0$, then $h$ must be regular at 
this point and \refs{tq} requires 
\eqnl{\beta_i= 4r_i h(a_i)-4r_i
\Bigl(\frac{f'}{f}-\frac{r_i}{z-a_i}\Bigr)\vert_{z=a_i}.}{bx}
To summarize, unless $r_i \es \ha$ for all $i$, 
the superconformal Ward-identity 
which requires $u$ to be a $U(1)$-section, leads to extra relations between 
the $a_i$, $\alpha_i$  
and the accessory parameters $\beta_i$ of $Q$. 
On the other hand we know from the theory of Fuchsian maps that there is 
only a discrete set of parameters $(\al_i,\beta_i,a_i)$ for which the 
upper half 
plane $H$ is mapped into an $\PSL(2,{\bf Z})$ polygon. Furthermore, for a 
given polygon these parameters are uniquely fixed once the origin and the 
scale in $H$ are chosen. We therefore conclude that \pan
{\it Unless all the angles $\al_i$ are zero the system is over-determined by 
the requirement that $u$ be a $U(1)$-section under equivalence 
transformations.} 

Unfortunately the accessory parameters $\beta_i$ of $Q$ are 
known 
explicitly only for some simple polygons \cite{Nehari}. This forces us to 
proceed in a somewhat roundabout way: For $\mu\es1$ the only polygon is the 
first domain in Fig. $1$. The Schwarzian for this polygon is given by 
\eqnl{Q(z) =
\frac{2}{9(z-1)^2} + \frac{2}{9(z+1)^2} - \frac{7}{36(z^2-1)}.}{QPLS}
Recalling that $T\in {\cal G}$, one finds that this 
polygon is compatible with a $f$-section only 
for $N_F\es 3$ in which case 
\eqnl{f(z) \propto (z^2-1)^{3/4}\mtx{and} \frac{\d u}{\d z} 
\propto 
\frac{36z^2 -31}{9(z^2 -1)^{\frac{1}{2}}}.}{duPSL} 
>From \refs{bbq} and \refs{qt} we obtain $\beta_1 = - \beta_2 = - 11/12$ 
which is incompatible with \refs{QPLS}. 
Alternatively one can directly integrate \refs{duPSL} to:
\eqnl{u(z) \propto 2\,z\,{\sqrt{{z^2} -1}} - \frac{13}{9} \,\log (z + 
{\sqrt{{z^2} - 1}}).}
{uPSL}
The presence of the log term shows that $u$ is not a $U(1)$-section.
Therefore $\mu=1$ never corresponds to a solution.\pan
Explicit inspection 
of all possible polygons with 
$2\leq\mu\leq6$ and their corresponding $f$-sections, taking into 
account the identifications due to $T\in{\cal{G}}$ and the boundary 
conditions, reveals that the exponents always satisfy $r_i\es\ha(1-\al_i)$ 
with exceptions only for $\al_i\es 0$. 
At these points, \refs{bbq} 
becomes simply 
\eqnl{h_i=\ha(r_i-\ha).}{bbq1}
If $n_0$ is the number of zero angles of 
the polygon $D$, 
the total residue condition \refs{cc} together with 
\refs{index} implies 
\eqnl{\frac{\mu}{6} - \frac{1}{4-N_F} = \frac{n_0}{2} 
- \sum_{i=1}^{n_0} r_i.}{suzero}
Suppose now that $r_i \neq \ha$ for all zero angles. 
Substituting \refs{bbq1} 
and \refs{at} into \refs{suzero} we then find $\mu=0$
which is a contradiction. We therefore conclude that there is at least one 
zero 
angle for which $r\es \ha$. In addition it follows from \refs{suzero} 
that for $\mu< 6/4-N_F$, any admissible polygon has at least two zero angles. 
We finish this section noting from \refs{tq} 
that $Q$, which carries the underlying $S\!L(2,\bbbz)$ structure 
can be constructed from the two $U(1)$-sections $f$ and $u$.

\section{Solution of the Model}
\label{sec5}
We consider the 
cases $N_F\es 1,2,3$ individually.\pan
\begin{center}
{\boldmath $N_F\es1$}
\end{center}
\pan
From \refs{bm} we have $\mu\leq 2$. 
Since $\mu=1$ has been excluded in the previous section, the only possibility 
is $\mu=2$ which corresponds to  
a double covering of the 
fundamental domain $D_0$ of $\PSL(2,\bbbz)$ (see Fig. $1$).
\iffigs

\begin{figure}[h]\caption{
Domains for $\mu=1$ and $2$
}
\input epsf
\epsfxsize=2.5cm
\begin{center}
\begin{tabular}{cc}
\epsffile{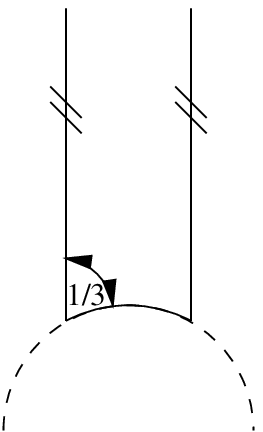}\hspace{3cm}&\hspace{3cm}\epsffile{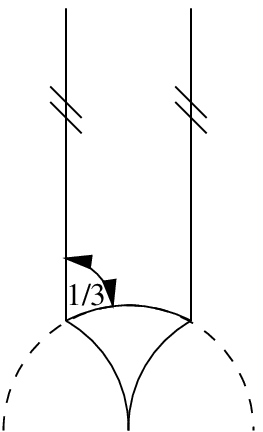 }
\end{tabular}
\end{center}
\end{figure}
\vspace{1cm}

\else
\message{No figures will be included. See TeX file for more
information.}
\fi
\noindent
This domain 
has angles $\alpha_1= 0$ 
and $\alpha_2=\alpha_3 = {2\over 3}$. The Schwarzian for this 
polygon is easily found to be 
\eqnl{Q(z) = \frac{1}{4z^2} + \frac{5}{36 (z-1)^2} + \frac{5}{36 (z+1)^2} 
- \frac{5}{18  (z^2-1)}.}{qsl2z}
There is a unique section $f$ compatible 
with \refs{cc} and \refs{fmc}. It is given by 
\eqnl{f(z)=f_0z^{1/2}(z^2-1)^{1/6},}{s2}
where $f_0$ is a constant. 
The equivalence group of this solution is a 
subgroup of index $2$ of $U(1) \times \PSL(2,{\bf Z})$.
Its elements 
contain an even number of $S$-generators of $\PSL(2,{\bf Z})$. Recall that 
the representation of ${\cal{G}}$ on $\tau$ is by a subgroup of
$\PSL(2,{\bf Z})$ for which the relation $(TS)^3\es1$ holds. Therefore the 
representation on $\tau$ of ${\cal{G}}$ generates all of $\PSL(2,{\bf Z})$. 
This explains the question raised in \cite{Nahm} concerning this solution. 
It also shows why it is important to define the equivalence group by its 
action on ${\bf a}$ rather than $\tau$.\pan
Using \refs{qsl2z}, \refs{s2} and integrating \refs{us2} 
we then have $u - u_0 \propto (z^2-1)^{\frac{1}{3}}$ where $u_0$ is a 
constant of integration. 
Proceeding as explained in \cite{USW1} and in the appendix, 
one then easily shows that for $u_0\es0$, ${\bf a}$ satisfies precisely 
the differential equation corresponding to the Seiberg-Witten solution 
\cite{Ito-Yang}.
Note that there is an ambiguity in the relation between $u$    
and $z$ due to the integration constant $u_0$. 
It is constrained to zero, either by an explicit $2$-instanton computation 
\cite{Dorey,Ito,Aoyama,Harano} or by imposing invariance under the discrete 
$R$-transformations $\phi\rightarrow e^{2\pi i/3}\phi$ \cite{SW2}.
\begin{center}
{\boldmath $N_F\es2$}
\end{center}
\pan 
This case is identical to $N_F\es0$ once the normalisation are chosen 
appropriately \cite{USW1}. We have $\mu\leq 3$.  
From the general discussion presented above, the two domains with $\mu\es1,2$ 
are however excluded  as they do not have enough zero angles. 
For $\mu\es3$ the only polygons $D$ are fundamental 
domains for $\Gamma_0(2)$ and without restricting the generality we can 
choose one 
which has only zero angles \cite{USW1}. 
Hence the necessary condition 
\refs{nec} is at the same time sufficient and therefore the 
superconformal anomaly imposes no further constraints. 
Consequently the unique solution for $N_F\es2$ has $\mu\es3$. 
The equivalence group is $\Gamma_0(2)$. Again this solution is precisely 
that proposed by Seiberg and Witten 
\cite{SW2}. The proof of this is identical to that presented in section $5$ 
of \cite{USW1}. The above analysis also shows that the more general 
definition of the equivalence group adopted here does not lead to new 
solutions for $N_F\es0$, as anticipated in the introduction. 
\begin{center}
{\boldmath $N_F\es3$}
\end{center}
\pan
This case is more involved. $\mu=1$ and $2$ are 
excluded from the previous discussion. $\mu=3$ can be shown to be 
incompatible with \refs{suzero}. For $\mu=4$ and $5$ 
we found one and three admissible polygons with $2$ zero 
angles  respectively. For $\mu\es6$ there is a 
whole set of polygons among which there is 
precisely one with only zero angles 
($n\es 3$), and which is a fundamental domain of $\Gamma_0(4)$. In that case 
again, the superconformal Ward-identity implies no new constraint and 
therefore this last polygon is a solution. This polygon has for 
Schwarzian $Q$,
\eqnl{Q(z)= \frac{1}{4 z^2} + \frac{1}{4(z-1)^2} + \frac{1}{4(z+1)^2} -
\frac{1}{2 (z^2-1)},}{s4}
and a unique $U(1)$-section $f$
\eqnl{f(z) \propto z^{1/2}(z^2-1)^{1/2}.}{f4}
Proceeding as before we find $u'\es\hbox{const}$. Again this solution is identical to the solution proposed 
by Seiberg and Witten modulo the undetermined 
integration constant in the relation between $u$ and $z$ whose 
value can only be fixed by explicit $2$-instanton computations 
\cite{Harano}.\pan
%
For all the other candidates  
the Ward-identity does lead to extra constraints and the system is therefore 
over-determined. We therefore expect these to be ruled out. However we are not 
able, presently to show this explicitly. Indeed, although for all these 
candidates we can write down a closed form for the Schwarzian $Q$, we have 
been unable to decide whether $Q$ corresponds to an $\PSL(2,{\bf Z})$-map, 
due to the lack of information about the accessory parameters $\beta_i$ in 
the Fuchsian maps \cite{Nehari}.  
\paragraph{Remark:} 
At this point a remark about possible candidates for alternative solutions 
in the framework of \cite{SW2} is of order. In \cite{D2} it was noted that 
in addition to the solutions proposed in \cite{SW2} the curves 
\eqnl{y^2= x^2(x-u)-\Lambda_2^4(x-cu)\mtx{for} N_F=2,}{anf2}
where $c\es1/9$, and 
\eqnl{y^2= x^2(x-u)-\Lambda_3^2u^2\mtx{for} N_F=3}{anf3}
satisfy the conditions $i)$ to $iv)$ given in section $11.3$ of 
\cite{SW2}. A careful analysis shows however that the curves 
\refs{anf2} and \refs{anf3} have unstable singularities (i.e. 
three roots coincide at some singular points in the $u$-plane). 
A possibly related property of these curves is that the holomorphic $1$-form 
$\lambda$ defined by
\eqnl{\pa_u\lambda=\frac{\sqrt{2}}{8\pi}\frac{dx}{y}+dw,}{lam}
where $w$ is a mereomorphic function, has non-vanishing residues. 
Indeed integrating \refs{lam} we get 
\eqngrl{\lam&=&2u\frac{dx}{y}+2\frac{x^2-\Lambda_2^4}{\Lambda_2^4c-x^2}
\frac{xdx}{y}\mtx{for} N_F=2,}
{\lam&=&6\frac{\Lambda_3^2u}{4\Lambda_3^2+
x}\frac{dx}{y}+\frac{2u-x}{4\Lambda_3^2+x}\frac{xdx}{y}
\mtx{for} N_F=3.}{Ilam}
Residues are however incompatible with two-dimensional monodromies. 
We conclude therefore that the curves (\ref{anf2},\ref{anf3}) do not lead 
to new solutions.

\section{Conclusions} 
Using supersymmetry, asymptotic freedom and assuming a finite number of 
strong-coupling singularities we have derived 
the low energy effective Lagrangians for, $N\es2$ supersymmetric 
$QCD$ with $N_F<3$ massless hypermultiplets. For $N_F\es3$ the argument 
is only almost complete. The problem however is not of a principle kind, but 
due to our lack of information about the accessory parameters for Fuchsian 
maps. An important role is played by the superconformal anomaly. 
It puts non-trivial constraints on an otherwise degenerate system in such a 
way that the solution becomes unique.\pan
The low energy Lagrangians we found are (up to a constant of integration) 
identical with those of Seiberg and Witten \cite{SW2}. It should be stressed 
however that we do not make any assumption concerning the role of the 
monopoles as in \cite{SW2} nor do we assume that the superconformal anomaly 
parameterizes the moduli space \cite{Matone2}. Rather we 
have shown that these two properties are consequences of more fundamental 
properties namely the extended supersymmetry and 
the underlying $\PSL(2,{\bf Z})$-structure of the theory. Although the 
finite mass 
condition, which reflects the asymptotic freedom of the underlying 
non-abelian theory, was used in the present derivation it is conceivable 
that this is itself a consequence of the superconformal Ward-identity. A 
more detailed analysis of the superconformal Ward-identity along the lines 
presented in section $4.2$ might answer this question. 

Our results also clarify the 'exceptional' role of $N_F\es1$ 
observed in \cite{Nahm}.

An interesting technical observation is that for cases discussed so far the 
upper bound for the index $\mu$ is always saturated. We believe that the 
reason for this originates in the condition that 
the superconformal anomaly be a 
$U(1)$-section,  but we have not been able to produce a general argument for 
this conjecture so far.

\vspace{1.5cm}

Our analysis of the $BPS$ formula was motivated by the constructive 
criticism of the referee. We would like to thank him for this. 
This work was started while I.S. visited the Isaac Newton Institute, 
Cambridge which is kindly acknowledged for its hospitality. Helpful 
discussions with O. Schnetz are also acknowledged.

\apendix

\setcounter{equation}{0}
\section{Differential Equations}

The simplest way to exhibit the equivalence between the solutions 
obtained in section \ref{sec5} and the corresponding 
Seiberg-Witten solutions is to write down their differential equations. 
First we fix the 
multiplicative 
constants appearing in $f$ and $\bf y$ and then obtain 
the differential equation satisfied by $\bf a$.
Without restricting the generality we may take $\bf y$ as 
in \refs{assy2}; the other constant $f_0$ introduced in 
\refs{fg2} is then determined as follows. 
The perturbative result for $\tau(a)$ has been derived in section  $2.1$ of 
\cite{D2}:
\eqnl{\tau (a) \simeq  \frac{i}{\pi} \log (\frac{a^{4-N_F}}{a_0}), \quad a_0 
= 2^{\frac{N_F-12}{2}} \Lambda_{N_F}^{4-N_F}}{a0} 
where $\Lambda_{N_F}$ is the dynamical scale in the 
Pauli-Villars scheme.
On the other hand, $\tau(z) \simeq \frac{i}{\pi} \log (\tau_0 z)$ where $\tau_0$ depends on $N_F$. Thus, 
\eqnl{a(z) \simeq  (a_0 \tau_0 z)^{\frac{1}{4-N_F}}.}{az}
It follows from \refs{fy}, \refs{assy2} and \refs{az} that  
$f_0 = (4-N_F) (a_0 \tau_0)^{\frac{1}{4-N_F}}$.\par
Since the procedure is the same in all cases we write explicitly 
the differential equation satisfied by $a$ only 
for $N_F\es 1$. 
$Q$ and $f$ are then given by \refs{qsl2z} and \refs{s2} respectively.
The constant $\tau_0$ is then determined noting that 
$z$ is related to the $\PSL(2,\bbbz)$-modular function $J$ by 
$J = - \frac{(z^2-1)^2}{4z^2}$ 
and by using the asymptotic limit for $\tau(J)$ 
\cite{Brandhuber}. 
With this we then obtain from 
\refs{fy}, \refs{ty} and \refs{us2} the 
 sought differential equation for ${\bf a}$: 
\eqnl{{\bf a}_{uu} + \frac{1}{4} \frac{u-u_0}{(u-u_0)^3 + \frac{27 
\Lambda_1^6}{16}}{\bf a} =0.}{auu1} 
Eqn. \refs{auu1} is precisely the differential equation satisfied 
by the Seiberg-Witten Ansatz \cite{Ito-Yang} provided that $u_0=0$ and that 
the scale 
$\Lambda_1^{\scriptscriptstyle \rm SW }$ used by Seiberg and Witten is 
related to the 
Pauli-Villars scale by $(\Lambda_1^{\scriptscriptstyle \rm SW } )^{3} = 4 
\Lambda_1^3$. 
In particular we recover in this way the result of \cite{D2,Pouliot}. 
The other 
cases $N_F=0,2$ and $3$ can be done in 
the same way. \par


\begin{thebibliography}{999}

\bibitem{SW1} N. Seiberg and E. Witten, Nucl. Phys. {\bf B426} (1994) 19,
(E) {\bf B430} (1994) 485.
\bibitem{SW2} N. Seiberg and E. Witten, Nucl. Phys. {\bf B431} (1994) 484.
\bibitem{Lerche} A. Klemm et al., Phys. Lett. {\bf B344} (1995) 169, 
Int. Jour. Mod. Phys. {\bf A11} (1996) 1929; P.C. Argyres and A.E. Faraggi, 
Phys. Rev. Lett. {\bf 74} (1995) 3931; P.C. Argyres and M.R. Douglas, 
Nucl. Phys. {\bf B448} (1995) 93; U.H. Danielsson and B. Sundborg, 
Phys. Lett. {\bf B358} (1995) 273.
\bibitem{Matone2} G. Bonelli, M. Matone and M. Tonin, Phys. Rev. 
{\bf D55} (1997) 6466, hep-th/9610026.
\bibitem{USW1} R. Flume, M. Magro, L.O'Raifeartaigh, I. Sachs 
and O. Schnetz, Nucl. Phys. {\bf B494} (1997) 331, hep-th/9611123. 
\bibitem{WH1} P.S. Howe and P.C.  West, Nucl. Phys. {\bf B486} (1997) 425, 
hep-th/9607239.
\bibitem{HW2} P.S. Howe and P.C.  West, Phys. Lett. {\bf B400} (1997) 30, 
hep-th/9611075.
\bibitem{Pouliot} D. Finnell and P. Pouliot, Nucl. Phys. {\bf B453} (1995) 225.
\bibitem{Dorey} N. Dorey, V.V. Khoze and M.P. Mattis, 
Phys. Lett. {\bf B388} (1996) 324; Phys. Rev. {\bf D54} (1996) 2921;
Phys. Rev. {\bf D54} (1996) 7832. 
\bibitem{Ito} K. Ito and N. Sasakura, Phys. Lett. {\bf B382} (1996) 95; 
Nucl. Phys. {\bf B484} (1997) 141;   
A. Yung, Nucl. Phys. {\bf B485} (1997) 38.
\bibitem{Aoyama} H. Aoyama, T. Harano, M. Sato and S. Wada, 
Phys. Lett. {\bf B388} (1996) 331.
\bibitem{Harano} T. Harano and M. Sato, Nucl. Phys. {\bf B484} (1997) 167.
\bibitem{Bilal} A. Bilal, hep-th/9601007. 
\bibitem{Ito-Yang} K. Ito and S.K. Yang, Phys. Lett. {\bf B366} (1996) 165; 
see also S. Ryang, Phys. Lett. {\bf B365} (1996) 113.
\bibitem{Sei} N. Seiberg, Phys. Lett. {\bf B206} (1988) 75.
\bibitem{AP} B. deWit, M.T. Grisaru and M. Ro\u cek, 
Phys. Lett. {\bf B374} (1996) 
297; A. Pickering and P. West, Phys. Lett. {\bf B383} (1996) 54.
\bibitem{OW} D. Olive and E. Witten, Phys. Lett. {\bf B78} (1978) 97.
\bibitem{Shif0} M.A. Shifman and A.I. Vainshtein, Nucl. Phys. 
{\bf B277} (1986) 456; A. Gorskii et al., Phys. Lett. {\bf B355} (1995) 466. 
\bibitem{W2} M.T. Grisaru, and W. Siegel, Nucl. Phys. {\bf B201} (1982) 292.
\bibitem{Rocek} G. Chalmers, M. Ro\u cek and R. von Unge, hep-th/9612195. 
\bibitem{MRSZ}  M. Magro, L.O'Raifeartaigh and I. Sachs, to appear.
\bibitem{Nehari} Z. Nehari, {\it Conformal Mapping}, 
1st Edition, McGraw-Hill, N.Y. (1952).
\bibitem{Venkov} A.B. Venkov, Functional Anal. Appl. {\bf 17} (1983), 
165.
\bibitem{Frakas} H. Frakas and I. Kra, {\it Riemann Surfaces}, 
Springer-Verlag, N.Y. (1980).
\bibitem{RS} S.J. Gates, M.T. Grisaru, M. Ro\u cek and W. Siegel, {\it 
Superspace; One thousand and one lessons in superspace}, 
Benjamin/Cummings (1983). 
\bibitem{CL} T.E. Clar, O. Piguet and K. Sibold, Nucl. Phys. {\bf B143} 
(1978) 445.
\bibitem{GW} M.T Grisaru and P.C.  West, Nucl. Phys. {\bf B254} 
(1985) 249.
\bibitem{Dorey2} N. Dorey, V.V. Khoze and M.P. Mattis, 
Phys. Lett {\bf B390} (1997) 205, hep-th/9606199; 
F. Fucito and G. Travaglini, Phys. Rev. {\bf D55} (1997) 1099, hep-th/9605215.
\bibitem{Matone} M. Matone, Phys. Lett. {\bf B357} (1995) 342.
\bibitem{Nahm} W. Nahm, hep-th/9608121.

\bibitem{D2} N. Dorey, V.V. Khoze and M.P. Mattis, Nucl. Phys. 
{\bf B492} (1997) 607, hep-th/9611016.
\bibitem{Brandhuber} A. Erd\'elyi et al., {\it Higher 
Transcendental Functions}, Vol. 3, McGraw-Hill, N.Y., (1955); 
A. Brandhuber and S. Stieberger, hep-th/9609130. 


\end{thebibliography}
\end{document}

\end